\documentclass[prl,aps,amssymb,amsfonts,amsmath,showpacs,preprint,superscriptaddress]{revtex4-1} 
\usepackage{graphicx} 
\usepackage{amsmath} 
\usepackage{amsfonts}

\newcommand{\ethaffil}{Laboratory of Physical Chemistry, ETH Zurich, 8093 Zurich, Switzerland}
\newcommand{\mplaffil}{New address: Max Planck Institute for the Science of Light, 91058 Erlangen, Germany}
\newcommand{\iclaffil}{New address: Department of Physics, Imperial College, London, United Kingdom}

\begin{document} \title{Controlling the phase of a light beam with a single molecule } 
\author{M. Pototschnig}
\affiliation{\ethaffil}
\author{Y. Chassagneux}
\affiliation{\ethaffil}
\author{J. Hwang}
\affiliation{\ethaffil}\affiliation{\iclaffil}
\author{G. Zumofen}
\affiliation{\ethaffil}
\author{A. Renn}
\affiliation{\ethaffil}
\author{V. Sandoghdar}
\affiliation{\ethaffil}\affiliation{\mplaffil} \email{vahid.sandoghdar@mpl.mpg.de}

\begin{abstract} We employ heterodyne interferometry to investigate the effect of a single organic molecule on the phase of a propagating laser beam. We report on the first phase-contrast images of individual molecules and demonstrate a
single-molecule electro-optical phase switch by applying a voltage to the microelectrodes embedded in the sample. Our results may find applications in single-molecule holography, fast
optical coherent signal processing, and single-emitter quantum operations. \end{abstract}

\maketitle
Single optical emitters such as atoms, ions, molecules, quantum dots and color centers have been most commonly detected via their fluorescence emission, which can be easily distinguished from the background with the help of spectral filters. Being
dependent solely on the population of the excited state, however, the fluorescence signal fails to access the coherent features of the interaction between the emitter and the incident
light. To examine both the amplitude and the phase involved in the coupling, one requires interferometric measurements. Until recently, the general wisdom was that the effect of a single
emitter in one arm of an interferometer would be too weak for direct measurements and, thus, high-finesse microcavities were used for such studies~\cite{Turchette:1995, Fushman:2008}. Interestingly,
laboratory efforts in the past half a decade have demonstrated more than 10\% attenuation of a laser beam by single organic molecules~\cite{Gerhardt:07a,Wrigge:08}, semiconductor quantum dots~\cite{Vamivakas:07}, and
trapped atoms~\cite{Tey:08}. In this process, the light that is coherently scattered from the emitter interferes with the
incident beam on the detector~\cite{Plakhotnik:01,Karrai:03,Gerhardt:07a,Wrigge:08}. Formulated in the framework of quantized light, the probabilities for the photon to be scattered by the
emitter and for going straight to the detector interfere. This interaction not only changes the amplitude but also affects the phase of the light beam~\cite{Zumofen:08,Zumofen:2009}.
Indeed, a phase shift of nearly one degree was recently measured on a weak laser beam using a single trapped atom in ultrahigh vacuum~\cite{Aljunid:09}. In this Letter, we report
on a phase shift greater than three degrees affected on a laser beam by a single molecule embedded in a solid. Furthermore, we use this effect to record the first phase-contrast images of
single molecules and to demonstrate a single-molecule electro-optical phase switch.

\begin{figure}[b!] \centering \includegraphics[width=7cm]{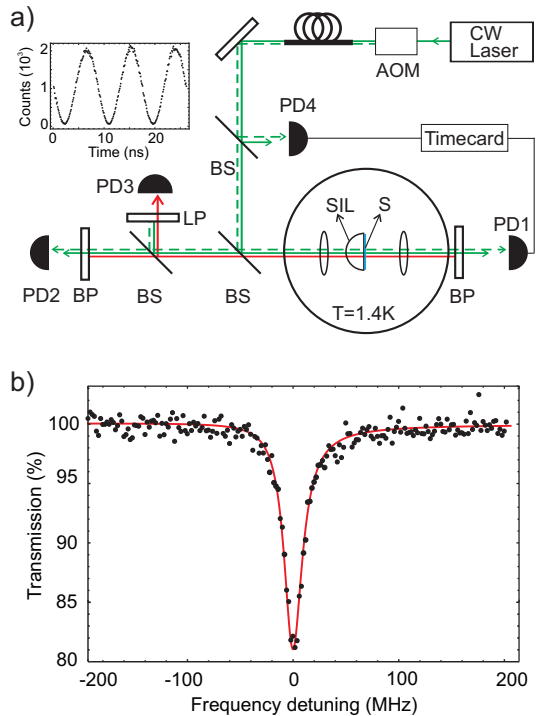} \caption{a) The experimental setup. BS: beam splitter; BP: bandpass filter; LP: low-pass filter; S: sample;
SIL: solid-immersion lens; AOM: acousto-optical modulator; PD: photodetector. The inset exemplifies raw data of the beating signal recorded in a start-stop
configuration using the signals of PD1 and PD4, respectively. Green solid and dashed lines indicate the two detuned laser beams after the AOM. The red solid line signifies the fluorescence signal from the sample. b) Laser beam attenuation of 18\% by a single molecule recorded on PD2 in reflection.} \label{setup}
\end{figure}

It is well established that when a harmonic oscillator is driven on resonance, its phase lags by $\pi/2$ radians. If the oscillator is a two-level dipolar scatterer driven by
a focused laser beam, this phase shift conspires with the Gouy phase of the incident beam to cause a $\pi$ phase difference between the incident and scattered components in the far field, leading
to an absorptive signal. The phase of the laser beam after interaction with the emitter is given by~\cite{Zumofen:2009,EPAPS} \begin{equation}
\varphi=\textrm{arg}\left(1-\eta e^{-i\psi}\frac{\Gamma^2+2 i \Delta \Gamma}{2\Omega^2+4\Delta^2+\Gamma^2}\right). \end{equation} Here, arg denotes the argument of its complex variable,
$\Gamma$ signifies the full width at half-maximum of the molecular resonance, $\Delta=\omega_L-\omega_0$ stands for the detuning between the frequencies of the laser ($\omega_L
$)
and the molecular transition ($\omega_0$), $\Omega$ is the Rabi frequency, and $\psi$ represents any accumulated geometrical phase difference between the scattered and transmitted fields in addition to the Gouy phase of $-\pi/2$. The parameter $0\leq\eta\leq1$ represents the product of geometrical and spectral factors of the
molecule and provides a measure for the efficiency with which it affects the laser beam. The
geometrical part of $\eta$ can reach unity in the ideal case where the incident beam and the molecular emission pattern are mode matched \cite{Zumofen:08}. The spectral contribution
to $\eta$ is unity for a two-level system, but it is reduced in molecules due to a finite branching ratio dictated by the Franck-Condon and Debye-Waller factors. In the work
presented here, we used the molecule dibenzanthanthrene (DBATT) embedded in a \emph{n}-tetradecane matrix, for which this contribution lies in the range between 0.2 and 0.5. 

In the weak excitation regime ($\Omega\ll\Gamma$), the maximal phase shift
of $\varphi=\arcsin\left(\frac{\mp\eta}{2-\eta}\right)$ is achieved at frequency detunings $\Delta=\pm \frac{1}{2} \Gamma \sqrt{1-\eta}$. As $\eta\rightarrow1$, this quantity
approaches $\pm \pi/2$ and moves to the resonance. In that situation, the intensity of the transmitted beam vanishes and the light is perfectly reflected~\cite{Zumofen:08}. For small values of $\eta$, the frequency
difference between the positive and negative extrema of the phase shift is close to $\Gamma$.

The experimental setup is sketched in Fig.~\ref{setup}a. The details of the sample preparation as well as the optical and mechanical features of the setup can be found in
Refs.~\cite{Wrigge:08,Hwang:09}. In short, the beam of a tunable ring dye laser is focused on a thin sample of \emph{n}-tetradecane that is doped with DBATT and prepared on a
solid-immersion lens. The transmitted and reflected lights are collected on avalanche photodiodes PD1 and PD2, respectively. As the frequency of the laser is scanned, individual
DBATT molecules in the illumination area become resonant and imprint their signatures onto the reflected and transmitted beams~\cite{Wrigge:08}. A noteworthy feature of our current
experimental configuration is that we integrated gold microelectrodes on the sample substrate to tune the resonance frequency of a molecule via the Stark effect.
Furthermore, by retro-reflecting the light from the electrode, we realized a double-pass excitation and achieved larger extinction dips (see Fig.~\ref{setup}b) beyond the
previously reported values of about 10\%~\cite{Vamivakas:07,Wrigge:08, Tey:08}.

To measure the change in the phase of the laser beam, we implemented a heterodyne interferometer as displayed schematically in Fig.~\ref{setup}a. The beam was first passed 
through
an acousto-optical modulator (AOM) to produce a second beam at a frequency detuning of 114.3 MHz. The two laser beams were equalized in power, coupled into a single-mode
optical fiber for perfect mode matching and then sent to a cryostat operating at T=1.4 K. The great advantage of this scheme is that due to a very small frequency
difference, the two laser beams experience essentially the same spatial and spectral effects in the entire optical path, making it very robust against mechanical, thermal, or
optical dispersion perturbations. We recorded the frequency beating between the two beams with a time-correlated single photon counting card in a start-stop configuration, using the TTL-converted signal of
a fast photodiode (PD4) as stop. The inset in Fig.~\ref{setup}a plots an example of the resulting temporal oscillations as a function of time delay. We reached visibilities above
95\%, mainly limited by the signal shot noise and the finite time response of the APD.

\begin{figure}[h!] \centering \includegraphics[width=7cm]{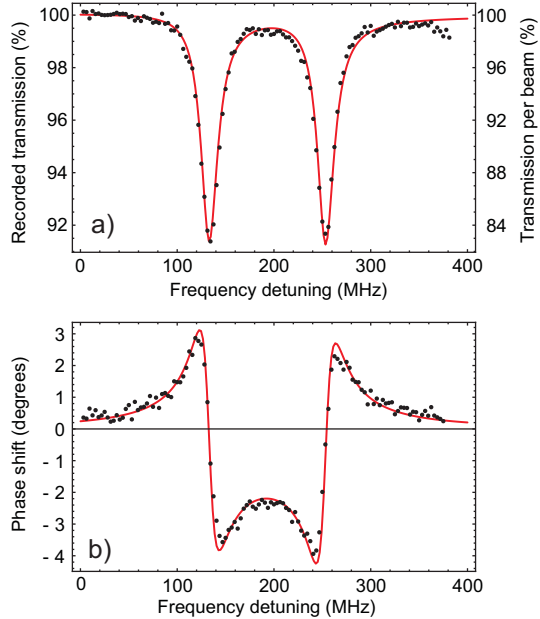} \caption{a) The signal on photodetector PD2 as a function of the detuning between the molecular resonance and the two laser
beams. The right-hand axis shows the corrected extinction dip corresponding to the effect of the molecule on one beam only. b)
The change of the phase recorded at the same time as the spectrum in (a). The solid curves display a simultaneous fit to (a) and (b).}\label{phase-scan} \end{figure}

As the laser frequency was swept across the resonance of the molecular transition, each of the two laser beams was affected by the presence of the molecule. 
Figure~\ref{phase-scan}a plots the resulting absorptive Lorentzian extinction spectrum with $\Gamma=21$ MHz measured on PD2. The dip size is reduced by a factor of two (see
Fig.~\ref{setup}b) because at each resonance only half of the total power seen by the detector interacts strongly with the molecule. The symbols in Fig.~\ref{phase-scan}b present
the measured phase shift imposed by the molecule on the laser beams as a function of the frequency detuning. We find that a single molecule
shifts the phase of each laser beam by about $\pm 3$ degrees close to its resonance. The slight asymmetry in Fig.~\ref{phase-scan}a and Fig.~\ref{phase-scan}b  results from a small axial displacement of the molecule from the focus and thus a slight deviation of $\psi$ from zero. 

\begin{figure}[t h!] \centering \includegraphics[width=8.5cm]{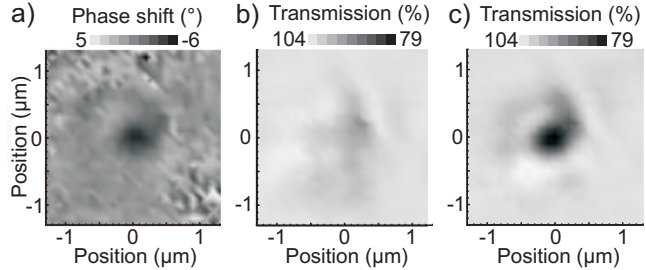} \caption{a) A phase-contrast image of a single molecule. The molecular resonance was
centered between the frequencies of the two laser beams (see Fig.~\ref{setup}). b) Simultaneously recorded transmission image. c) Image of the molecule extracted from the transmission signal on resonance.}\label{phase-image} \end{figure} 

One of the powerful applications of phase-sensitive measurements in optics has been in holography~\cite{Gabor:1949}, phase-contrast imaging~\cite{Zernike:1942a}, and differential interference contrast microscopy. An important advantage of the latter methods is that they visualize samples that do not show a notable contrast in absorption but undergo variations in the optical path. One trivial such case
occurs if an object contains pure topographical modulations. However, even for samples with material heterogeneity the signal-to-noise ratio (SNR) of a phase-contrast measurement turns out to be greater than its absorptive counterpart if material resonances are far detuned from the illumination wavelength. This is because the wings of a dispersive Lorentzian profile drop as $1/\Delta$ whereas those of its absorptive counterpart decay faster as $1/\Delta^2$~\cite{EPAPS}. We remark though, that the best SNRs obtained for the detection of an emitter via phase and intensity are comparable if one restricts the total number of the scattered photons as a meter for the destructiveness of the measurement process~\cite{Lye:03, EPAPS}. 

The sensitivity in our phase measurements implies that we should now be able to record phase contrast images of single molecules. Figure~\ref{phase-image}a shows such an image as the laser focal spot was scanned laterally across the sample. Here, the molecular resonance frequency was placed at the center of the two 
laser frequencies, i.e. at a detuning of about $\pm 57 \rm{MHz}\approx3\Gamma$. For comparison, in Fig.~\ref
{phase-image}b we present the extinction (transmission) image of the same region under identical conditions. It is evident that in this frequency-detuned case, the 
contrast in the extinction image is inferior to that of the phase measurement. On the other hand, Fig.~\ref{phase-image}c shows that a high-contrast extinction image can be recovered on resonance. Extension of these studies to the axial dimension would provide a three-dimensional holographic image of individual molecules in the sample.

\begin{figure}[h!] \centering \includegraphics[width=7.5cm]{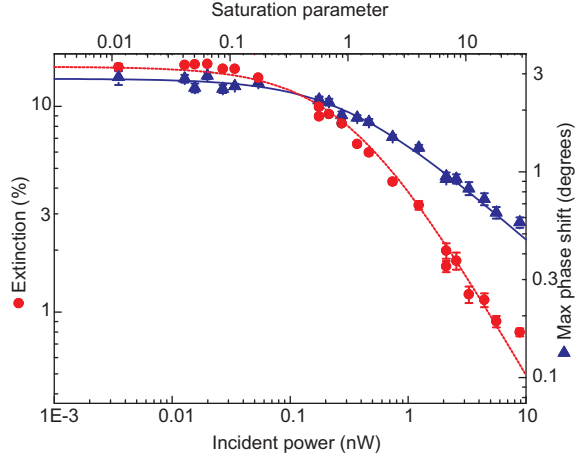} \caption{Extinction and phase signals versus excitation power.}\label{phase-power}
\end{figure}

Phase measurements report on the coherent elastic scattering component of resonance fluorescence, which diminishes at high excitation powers~\cite{cohen-book,Wrigge:08}. It is,
thus, instructive to investigate the effect of a single molecule on the phase of strong laser beams. Similar to its dependence on detuning, it follows from Eq.~(1) that the extinction signal vanishes as
$1/\Omega^2$ whereas the maximum phase shift scales as $1/\Omega$. In Fig.~\ref{phase-power}, the blue triangular symbols display the maximum of
the recorded phase shift over four decades of incident power. The red circles plot the results of the simultaneously measured attenuation signal. It turns out that the
phase shift signal is lost more slowly at high power levels than the extinction dip. 

We note that although the spectrum in Fig.~\ref{phase-scan}b could be described as the sum of
two independent dispersive Lorentzians, a close scrutiny of the data recorded at high excitation powers reveals clear deviations from this simple consideration due to several
effects. First, the AC-Stark shift and power broadening reduce the effective frequency difference between the two beams. Second, for very high powers a hyper-Raman effect caused by
the coherent cooperative interaction with the two laser beams becomes important~\cite{Lounis:1997}. In this process two photons at one laser frequency $\omega_{1}$ lead to the
excitation of the molecule and a scattered photon at the second laser frequency $\omega_{2}$ if the resonance condition $\omega_{2}=2\omega_1-(\omega_{0}+\delta')$ is met, 
where
$\delta'$ denotes the AC Stark shift. The third point has to do with the fact that higher harmonics of the difference frequencies $ (\omega_2 - \omega_1)$ contribute to the beating signal. Thus, to obtain a quantitative understanding of the lineshapes at high powers, we used Floquet theory ~\cite{Papademetriou:92, EPAPS}. We fitted the experimental spectra to extract the data for a single laser beam as shown in Fig.~\ref{phase-power}. The solid curves in this figure represent the theoretical behavior assuming $\eta=0.1$.

\begin{figure}[h!] \centering \includegraphics[width=8.5cm]{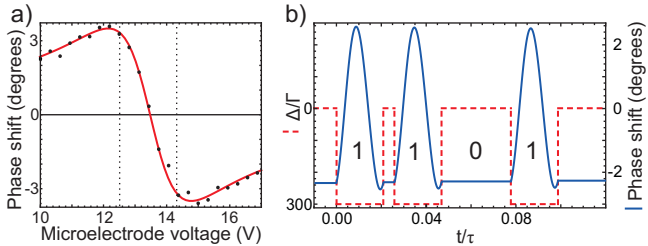} \caption{a) Measured phase shift of the excitation laser beam as a function of the voltage applied to
the microelectrodes in the sample. The dashed lines indicate the optimal switching range. b) Calculated phase shift (solid blue) when a series of rapid voltage pulses (dashed red) are applied.}\label{starkgating}
\end{figure}

The sharp slopes of the dispersive Lorentzian profiles in Fig.~\ref{phase-scan}b can be exploited for rapid phase switching and gating by varying the frequency detuning between a
laser beam and the molecule. To explore this possibility, we applied a voltage to the microelectrodes that were fabricated on the sample substrate at separations of about 20 $\mu m$ and shifted the molecular resonance via the Stark
effect. As shown in Fig.~\ref{starkgating}a, a potential change of less than 2V resulted in a phase shift of 6 degrees. This phase switching process can be faster than the spontaneous lifetime, which is of the order of nanoseconds for typical optical emitters. Figure~\ref{starkgating}b displays an example of calculations where four logic voltage pulses 1, 1, 0, 1 were applied to shift the molecular resonance from $+\Gamma$ to $-300\Gamma$. We find that the phase of the light beam is shifted by more than 4 degrees within $10^{-2}\tau$, where $\tau$ is the excited state lifetime. The performance details of this phase switch can be varied by determining the start and end values of the applied frequency detuning, but they follow the general guideline that the larger the detuning, the larger the generalized Rabi frequency and therefore the faster the phase reaction will be~\cite{EPAPS}. Considering that Stark shifts as large as $300\Gamma$ have already been realized using nanoelectrodes~\cite{Gerhardt:09b}, single molecules should be able to offer switching speeds of THz, only limited by the bandwidth of the electrodes. An experimental demonstration in this regime is left for future work as it requires faster detectors.

We have shown that a single organic molecule embedded in a solid matrix can shift the phase of a propagating light beam by several degrees. Improvements to the emitter-light coupling toward the ultimate limit of perfect reflection~\cite{Zumofen:08,Zumofen:2009} can make our scheme highly attractive for quantum gate operations~\cite{Nielsen-Chuang}. A particularly attractive on-chip solution in this regard is offered by the near-field coupling of molecules to light in metallic or dielectric nanoguides~\cite{Akimov:07, Quan:09}. This arrangement can be conveniently combined with integrated interferometers and provides the possibility to amplify the phase shift by coupling to several molecules placed in series within a few hundred nanometers. Such an approach is simpler, more compact and more robust than those based on microcavities~\cite{Turchette:1995, Fushman:2008, Faraon:2010} because the system occupies a much smaller volume and does not rely on stringent resonant conditions of high-finesse cavities. Extension of the concepts demonstrated in this work to emitters with $\Lambda$ or V energy level schemes would allow all-optical phase switching at the single-photon level. However, the strong optical nonlinearity of single molecules would also be helpful in the classical regime, where effects such as finite carrier lifetime and two-photon absorption under intense illumination currently limit the state-of-the-art phase switsches~\cite{Koos:2009}. 

This work was supported by the Swiss National Foundation (SNF) and the ETH Fellows program. We thank S. G\"otzinger for experimental support and
M. Krishnan for the preparation of the microelectrodes.

%

\end{document}